# Individual-Level SNP Diversity and Similarity Profiles

Zhanshan (Sam) Ma[12]*    Lianwei Li[1]    Ya-Ping Zhang[12]*

[1]State Key Lab of Genetic Resources and Evolution
Kunming Institute of Zoology
Chinese Academy of Sciences
[2]Center for Excellence in Animal Evolution and Genetics
Chinese Academy of Sciences
Kunming 650223 China
*For all correspondence: ma@mail.kiz.ac.cn or zhangyp@mail.kiz.ac.cn

**Abstract**—Classic concepts of genetic (gene) diversity (heterozygosity) such as Nei (1973: PNAS) and Nei & Li (1979: PNAS) nucleotide diversity were defined within the context of populations. Although variations are often measured in population context, the basic carriers of variation are individuals. Hence, measuring variations such as SNP of individual against a reference genome, which has been ignored currently, is certainly of its own right. Indeed, similar practice has been a tradition in ecology, where the basic framework of diversity measure is individual community sample. We propose to use Renyi's-entropy-derived Hill numbers to define SNP (single nucleotide polymorphism) diversity (including alpha-, beta-, and gamma-diversities) and similarity profiles. Hill numbers are derived from Renyi's entropy, of which Shannon's entropy is a special case and which have found widely applications including measuring the quantum information entanglement, wealth distribution in economics and ecological diversity. The newly proposed SNP diversity not only complements the existing genetic diversity concepts by offering individual-level metrics, but also offers building blocks for comparative genetic analysis at higher levels.  The profile concept also helps to resolve a dilemma in measuring diversity—the choice from various diversity indexes, because diversity profile unifies some of the most commonly used indexes (as special cases) with different diversity *orders* (along the rareness-commonness spectrum of gene mutations). Finally, the profiles can be estimated with rarefaction approach, which may help to relieve some effect of insufficient sequencing coverage.

# Introduction

SNPs (single nucleotide polymorphism) are single-nucleotide substitutions of one base for another and arguably the commonest genetic variation. There are two general categories of approaches to investigating SNPs: one is the genomic approach and another is the functional approach. With genomic approach, scientists have catalogued the SNP database in the 3-billion-base pair human genome (*e.g.,* https://www.ncbi.nlm.nih.gov/snp/, http://www.hgvs.org/central-mutation-snp-databases). The functional approaches have been adopted by scientists and clinicians who are interested in the implications of SNPs to a particular disease or drug response. With either approaches, statistically characterizing the *abundance* and *distribution* of SNPs is both important but challenging. Many existing characterizations of SNPs have been performed such as computing heritability (*e.g.*, Yang *et al*. 2017), computing gene and pathway scores to improve statistical power and gain biological insight (*e.g.*, Lamparter *et al*. 2016), genetic variation analysis (*e.g.*, The Genomes Project Consortium, 2015), and distribution fitting (*e.g.*, Tang *et al*. 2016).

Amos (2010) found that the distributions of even small SNP clusters are non-randomly distributed in the human genome. In other words, SNPs are not distributed at random across the chromosome or whole genome, but are aggregated or clustered. A variety of processes from ascertainment biases (*i.e.*, the preferential development of SNPs around interesting genes) to the action of mutation *hot spots* and natural selection may be responsible for the highly non-random distribution of SNPs. For example, natural selection may modulate local variability along a chromosome to generate non-randomness. The distribution of SNPs along a chromosome is frequently harnessed to infer the action of natural selection. The non-random distribution of SNPs has far reaching ramifications for how to statistically characterize SNPs properly, in particularly, the choice of summary statistics. For example, the non-random distribution makes many of the commonly used *aggregation* functions such as arithmetic mean (average) and even median poor metrics for characterizing SNPs (*e.g.*, Beliakov *et al*. 2016, James 2016). Instead, the entropy-based *aggregation functions* such as Shannon's entropy and Renyi's general entropy should be more appropriate for summarizing the information transpired by SNPs. In fact, Shannon entropy, which was borrowed from Shannon (1948) information theory, has been the most widely used metric for measuring species diversity (also known as ecological diversity, community diversity or biodiversity), although recent studies (Chao et al. 2012, 2014, Jost 2007, Ellison 2010) have reached a consensus that the Hill numbers, which are derived from Renyi's general entropy, offer the most appropriate alpha-diversity measures, and are advantageous for multiplicatively partitioning beta-diversity. In following sections, we will define the SNP diversity with Hill numbers and obtain a series of metrics for summarizing the distribution of SNPs.

Of course, measuring diversity with entropy is not new at all, and the concepts of genetic (gene) diversity (heterozygosity) have been proposed and widely applied since pioneering works in 1970s (Nei 1973, Nei & Li 1979). We observed that all existing genetic (gene) diversity have been defined within the context of populations. Although variations are often measured in population context, the basic carriers of variation are individuals. Hence, measuring variations such as SNP of individual against a reference genome, which has been ignored currently, is certainly of its own right. Indeed, similar practice has been the tradition in ecology, where the basic framework of diversity measure is individual community sample. We fill this gap in existing literature of genetic (gene) diversity by learning from ecology to define individual-level SNP diversity and similarity profiles.

In ecology, Hill numbers (Hill 1970) capture the essential properties of species abundance distribution (SAD) in a community and hence provide effective metrics for measuring species diversity because SAD contains full diversity information about a community. Hill numbers were derived from Renyi (1961) general entropy, of which Shannon entropy is a special case, and which has found wide applications in various fields of science and technology, from measuring quantum information entanglement to the wealth distribution in economics, and more recently from measuring ecological diversity (*e.g.*, Chao *et al* 2012, 2014) to measuring metagenome diversity (Ma & Li 2018). As reiterated in Sherwin et al. (2017), information theory has been playing a broadening role in molecular ecology and evolution. Similar their critical roles in measuring ecological diversity, Hill numbers can capture essential properties of the SNP distribution on a genetic entity such as a chromosome and offer effective metrics for measuring SNP diversity.

The primary objective of this article is to define the SNP diversity with Hill numbers at the individual level, including the alpha diversity, beta diversity, and gamma diversity of SNPs. Of course, to define SNP of an individual, a reference genome is required. Therefore, to define SNP diversity, two individuals including a reference genome and a target genome are required. In contrast, existing concepts (indexes) of genetic (gene) diversity were all defined in a population of more than two individuals. The SNP alpha-diversity we will define, in effect, measures the unevenness or heterogeneity of SNPs in a genetic entity such as a chromosome or genome at the individual level. This not only complements the current population-level genetic (gene) diversity, but also provides building blocks for further comparative SNP analyses. For example, our SNP beta-diversity is defined to measure the difference between two or more individuals, and SNP gamma diversity is defined to measure the total diversity within the individuals of a population. Therefore, our concept and supporting metrics of SNP diversity provide a cross-scale tool for analyzing SNP variations at both individual and population levels.

We also define four SNP similarity metrics based on the Hill numbers. The SNP similarity metrics can be utilized to directly compare the SNP distribution patterns of the so-termed *N*-population, *i.e.*, a population or cohort consisting of *N* individuals. Together, SNP diversity and similarity measures in Hill numbers offer effective tools to reveal genetic and evolutionary insights SNPs may reveal. We demonstrate the computations of SNP diversity and similarity measures with the SNP datasets obtained from whole-genome sequencing of 9 individuals, consisting of four lung cancer patients and their five healthy relatives (Kanwal *et al*. 2017).

As a side note, our title used the term "profile" (of diversity/similarity), which is to do with the definitions of Hill numbers. Hill numbers (also termed *diversity profile*) are a series of diversity measures that are weighted differently by the occurrences of low frequency SNPs, which form the long tail of the highly skewed SNP distribution and is often responsible for the biggest challenge in characterizing the SNP properly and effectively. Hence, the diversity/similarity profiles based on the Hill numbers are ideal for dealing with the challenge from the non-random distribution nature of human SNPs mentioned previously. The diversity profile also avoids a serious issue associated with most existing diversity indexes, *i.e.*, there was not a single diversity index that can comprehensively measure diversity but multiple indexes (such as Shannon and Simpson indexes) are not comparable with each other. This makes the choice of diversity index often confusing for practitioners: which one, Simpson's index or Shannon's index is better?

Before proceeding to propose and develop our individual-level SNP diversity, here we summarize the following four points to answer a possibly question from readers. Why bother to introduce another level of diversity even if it can be properly defined? (*i*) The SNP *alpha*-diversity profile offers a series of metrics for characterizing the SNP patterns of an individual genome, which is personal and individual-specific at the whole genome level. (*ii*) It also offers a cross-scale tool for comparing individuals and complements the population level analysis. For example, the SNP *beta*-diversity (we propose) is defined to compare two or individuals within a population in their SNP distribution variation patterns. SNP *gamma*-diversity (we proposed) is defined to measure the total diversity (variations) of all individuals within a population. (*iii*) The study also presents another example of the cross-fertilizing between population genetics and community ecology. (*iv*) In our opinion, the case for developing an individual level genetic diversity is particularly compelling in the genomics era when the genetic information of an individual in the form of DNA sequences is readily available, while in the 1970s, the data for individual-level is much small and only population data were big enough to require formal metrics. As a side note, our proposed metrics can also be applied to extended population-level genetic diversity, which we will address in a follow-up study.

## Concepts and Definitions

Let us start with a brief review on the species diversity (*aka* community diversity, biodiversity or ecological diversity) to explain the two essential elements of diversity concept in general, which should facilitate the introduction of our SNP diversity and similarity measures below. Species diversity refers to the ecological diversity of species in an ecological community, but diversity concept is equally applicable to genetic diversity (*e.g.*, Nei 1973, Wehenkel *et al*. 2006, Bergmann *et al*. 2013) or other entities such as metagenome diversity (Ma & Li 2018). Conceptually, diversity possesses two essential elements: the *variety* and the *variability* of *varieties* (Gaston 1996; Chao *et al*. 2014). For example, the two elements of species diversity are species (variety) and the variability of species abundances. To quantify the concept of species diversity, one surveys a community (usually by *sampling*), counts the abundances of each species in the community, and obtains $p_i$=(the relative abundance of species $i$)=(the number of individuals of species $i$)/(the total individuals of all species in the community), and also counts the number of species in the community ($S$). The dataset from such a survey (sampling) is a vector of species abundance in the form of ($p_1, p_2, …, p_i, …p_S$). For such a vector of relative abundances (frequencies), one approach to characterizing it is to fit a statistical distribution, which is known as species abundance distribution (SAD) in community ecology. The most widely used SADs include log-series, log-normal, and power law distributions; a common property of SADs is that they are highly skewed, long tail distributions, but rarely follow the normal distribution or Poisson distribution. The latter two are arguably two most commonly used statistical distributions in general biostatistics. Therefore, the SAD is highly aggregated (skewed or non-random), just as the non-random SNP distribution previously mentioned in the introduction section. Although SAD fully describes the species abundance frequency and therefore adequately captures the full characteristics of species diversity, using a SAD to measure diversity fails to present intuitive measures to synthesize the two elements of diversity (*i.e.*, variety and variability), and is therefore highly inefficient. An alternative approach to fitting SAD is to use various diversity metrics (also known as measures or indexes). Numerous diversity metrics for measuring species diversity have been proposed, with Shannon's entropy index being the most well known.

Diversity metrics belong to the so-termed *aggregate* functions, which combine several values into a single value (Beliakov *et al*. 2016, James 2016). The arithmetic mean (average) is the most commonly utilized aggregation function, but it is a rather poor metric for measuring diversity due to the highly non-random distribution of species abundances. Instead, entropy-based aggregation function is suitable for measuring diversity. The first and also still one of the most widely utilized entropy-based diversity metric is Shannon entropy index, which was attributed to Claude Shannon, the co-founder of information theory (Shannon 1948; Shannon & Weaver 1949), but Shannon had never studied biodiversity himself. What happened was that ecologists borrowed the idea from Shannon's information theory, in which Shannon's entropy measures the content of information or uncertainty in communication systems. Of course, Shannon's entropy is indeed sufficiently general for measuring biodiversity because diversity is essentially heterogeneity, and heterogeneity and uncertainty both can be measured by the change of information, *i.e.*, information lowers uncertainty.

Using Shannon entropy as example, species diversity ($H$), more accurately species evenness, can be computed with the following formula,

$$H = -\sum_{i=1}^{S} p_i \ln(p_i) \qquad (1)$$

where $S$ is the number of species in the community, and $p_i$ is the relative abundance of each species in the community. In terms of the *variety-variability* notion for defining diversity, the *variety* is the species and *variability* is the species abundance obviously. In fact, the variety-variability notion can be utilized to define diversity for any systems (not even limited to biological systems) that can be abstracted as the two elements of variety and variability, including SNP diversity, as exposed below.

## Definitions for SNP diversities

Using an analogy, a chromosome that has many *loci* is similar to an ecological community of many species, and each *locus* may have different number of SNPs. With variety-variability notion for defining diversity, the locus is the *variety* (similar to species in a community), and the number of SNPs at each locus is the *variability* (similar to species abundance in a community). Assuming $S$ is the number of *loci* with any SNP, and $p_i$ is the *relative* abundance of SNPs at locus $i$ (*i.e.*, the number or abundance of SNPs at locus $i$ divided by the total number of SNPs from all loci), then SNP diversity can be measured with Shannon entropy (Eqn. 1). Strictly speaking, SNP may also be termed *locus* diversity, since *locus* is essentially the 'habitat' where SNPs reside. Fig 1 conceptually illustrated the distribution of SNPs on a chromosome; specifically how $p_i$ is defined and computed.

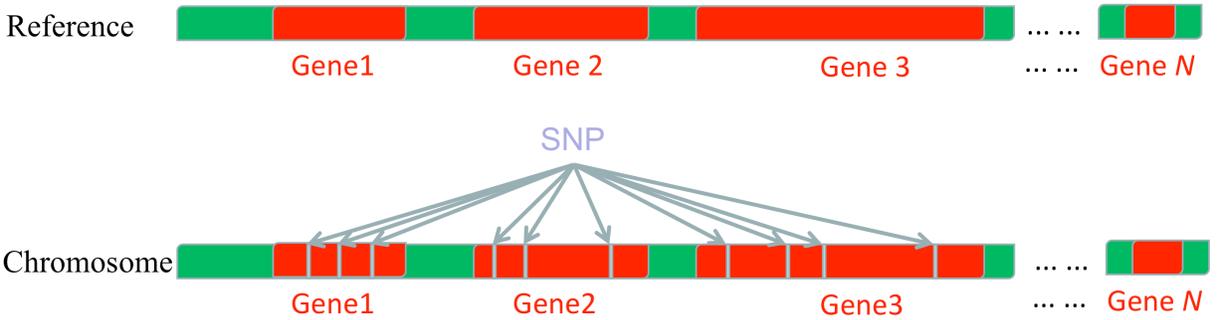

**Fig 1**. A conceptual diagram showing the distribution of SNPs on a chromosome with reference to the reference chromosome: the number of SNPs on a gene locus is similar to the species abundance in an ecological community. For example, there are three SNPs on the locus of gene-1, assuming the total SNPs on the chromosome is $N$ (or 10 displayed with the first 3 genes displayed), then the relative SNP abundance for gene-1 is equal to $3/N$ (or $3/10=0.3$ with the 3 genes displayed). Similarly, $p_2$, $p_3$, ... can be computed.

Although Shannon's entropy has been widely used for measuring species diversity, a recent consensus among ecologists is that Hill numbers, which are based on Renyi's general entropy, offer the most appropriate metrics for measuring alpha-diversity and for multiplicatively partitioning beta-diversity (Chao *et al*. 2012, 2014, Ellison 2010). Given the advantages of Hill numbers over other existing diversity indexes, we believe that the Hill numbers should also be a preferred choice for defining the SNP diversity.

### (*i*) SNP alpha-diversity
Hill numbers were derived by Hill (1973) based on Renyi's (1961) general entropy. Here we propose to apply it for defining the SNP alpha-diversity, *i.e.*,

$$^{q}D = \left(\sum_{i=1}^{G} p_i^q\right)^{1/(1-q)} \qquad (2)$$

where $G$ is the number of gene loci with any SNP, $p_i$ is the relative abundance (*i.e.*, the *frequency of occurrence*) of SNPs at locus $i$, $q=0, 1, 2, \ldots$ is the *order* number of SNP diversity, $^qD$ is the SNP alpha-diversity at diversity order $q$, *i.e.*, the Hill numbers of the $q$-th order.

The Hill number is undefined for $q=1$, but its limit as $q$ approaches to *1* exists in the following form:

$$^1D = \lim_{q \to 1} {}^qD = \exp\left(-\sum_{i=1}^{G} p_i \log(p_1)\right) \qquad (3)$$

The diversity order ($q$) determines the sensitivity of the Hill number to the relative abundance (*i.e.*, the frequency of occurrence) of SNP. When $q=0$, the SNP frequency does not count at all and $^0D=G$, *i.e.,* the *SNP richness,* similar to the *species richness* in species diversity concept. When $q=1$, $^1D$ equals the *exponential* of Shannon entropy, and is interpreted as the number of SNPs with typical or common frequencies. Hence, Shannon index is essentially a special case of Hill numbers at diversity order $q=1$. When $q=2$, $^2D$ equals the reciprocal of Simpson index, *i.e.*,

$$^2D = (1/\sum_{i=1}^{G} p_i^2) \qquad (4)$$

which is interpreted as the number of dominant or very frequently occurred SNPs. Therefore, two most widely used diversity indexes, Shannon index and Simpson index are the special cases of the Hill numbers.

In general, we need to specify an entity (unit or scope) for defining and measuring SNP diversity. For demonstrative purpose in this article, we choose individual chromosome as the entity for defining SNP diversity, similar to using community for defining species diversity. The general interpretation of diversity of order $q$ is that the chromosome contains $^qD=x$ loci with equal SNP frequency. Note that the entity for defining SNP diversity can be other appropriate units such as the *whole genome* of an organism or segment of chromosome.

The above-defined SNP diversity measures the diversity of SNP on an individual genetic entity (such as chromosome or genome), similar to the concept of alpha diversity in community species diversity, and we term it *SNP alpha-diversity*. In the following, we define the counterparts of species beta-diversity and gamma-diversity in community ecology for SNPs, *i.e.*, *SNP beta-diversity* and *SNP gamma-diversity*.

**(*ii*) *SNP Gamma Diversity***
While the previously defined SNP alpha-diversity is aimed to measure the SNP diversity within a genetic entity (such as a chromosome or genome), the following SNP gamma-diversity is defined to measure the *total* SNP diversity of pooled, multiple ($N$) chromosomes from a population (cohort) of $N$ different individuals, one from each individual but with the same chromosome numbering.

Assuming there are $N$ individuals in a population (cohort), we define the *SNP gamma-diversity* with the following formula, similar to the species gamma-diversity in ecology (*e.g.*, Chao *et al.* 2012, 2014; Chiu *et al.* 2014),

$$^qD_\gamma = \left(\sum_{i=1}^{G} (\overline{p_i})^q\right)^{1/(1-q)}, \qquad (5)$$

where $\overline{p_i}$ is the SNP frequency on the *i*-th locus ($i=1,2,\ldots,G$) in the pooled population of $N$ individuals (termed *N-population*).

Comparing Eqn. (5) for gamma diversity with Eqn. (2) for alpha diversity reveals that the gamma-diversity is the Hill numbers based on the SNP *frequency* at $i$-th locus in the $N$-population. Similar to Chao *et al*. (2012, 2014), Chiu *et al*. (2014) derivation for species gamma-diversity in ecological community, assuming $y_{ij}$ is the SNP frequency at $i$-th locus of $j$-th individual, $y_{i+}$ is the total value of SNP at $i$-th locus contained in the $N$ individuals, $y_{+j}$ is the total SNP from $j$-th individual, $y_{++}$ is the total SNP contained in $N$ individuals, $p_{ij}$ is the SNP frequency at $i$-th locus of $j$-th individual, $w_j$ is the weight of the $j$-th individual,

$$y_{i+} = \sum_{j=1}^{N} y_{ij} = yy_{++} \sum_{j=1}^{N} w_j p_{ij}$$

$$y_{+j} = \sum_{i=1}^{G} y_{ij}$$

$$y_{++} = \sum_{i=1}^{G} \sum_{j=1}^{N} y_{ij}$$

$$p_{ij} = y_{ij} / y_{+j}$$

$$w_j = y_{+j} / y_{++}, \quad \sum_{j=1}^{N} w_j = 1,$$

it can be easily derived that,

$$\overline{p}_i = (y_{i+}/y_{++}) = \sum_{j}^{N}(w_j p_{ij}). \tag{6}$$

Plug Eqn. (6) for $\overline{p}_i$ into the definition of *SNP gamma diversity* [Eqn. (5)], we obtain the following formulae for computing *SNP gamma-diversity* of $N$-population as follows:

$$^qD_\gamma = \left(\sum_{i=1}^{G}(\overline{p}_i)^q\right)^{1/(1-q)} = \left\{\sum_{i=1}^{G}\left(\frac{y_{i+}}{y_{++}}\right)^q\right\}^{1/(1-q)} = \left\{\sum_{i=1}^{G}\left(\sum_{j=1}^{N} w_j p_{ij}\right)^q\right\}^{1/(1-q)} \quad (q \neq 1) \tag{7}$$

$$^1D_\gamma = \lim_{q \to 1} {}^qD_\gamma = \exp\left\{-\sum_{i=1}^{G}\left(\frac{y_{i+}}{y_{++}}\right)\log\left(\frac{y_{i+}}{y_{++}}\right)\right\} = \exp\left\{-\sum_{i=1}^{G}\left(\sum_{j=1}^{N} w_j p_{ij}\right)\log\left(\sum_{j=1}^{N} w_j p_{ij}\right)\right\} \quad (q = 1) \tag{8}$$

### (*iii*) SNP beta diversity

In community ecology, there are two schemes for defining beta-diversity: one is the additive partition and another is the multiplicative partition of gamma diversity into assumingly independent alpha-diversity and beta-diversity. Recent consensus (*e.g.*, Jost 2007; Ellison 2010; Chao *et al*. 2012, 2014, Gotelli & Chao 2013; Gotelli & Ellison 2013) recommended the use of multiplicatively partitioned beta-diversity by partitioning gamma diversity into the product of alpha and beta diversities, in which both alpha ($^qD_\alpha$) and gamma ($^qD_\gamma$) diversities are measured with the Hill numbers. That is, beta-diversity is defined as:

$$^qD_\beta = {}^qD_\gamma / {}^qD_\alpha \tag{9}$$

We adopt the exactly same multiplicative partition of the Hill numbers in species diversity for measuring SNP beta-diversity except that both alpha- and gamma- diversities are computed with SNP frequency (relative abundance), rather than with species abundances.

This SNP beta-diversity ($^qD_\beta$) derived from the above multiplicative partition takes the value of *1* if all communities are identical, and the value of *N* (the number of individuals in the population) when all individuals are completely different from each other (*i.e.*, no shared SNPs).

Although Eqn. (2) correctly defines the SNP alpha-diversity, it requires some adaptations to apply for the partition of gamma diversity in order to obtain beta-diversity with Eqn. (9). Similar to the derivation for species alpha diversity as demonstrated in Chiu *et al*. (2014), we can derive the following formulae for SNP alpha diversity in *N*-population setting, *i.e.*,

$$^qD_\alpha = \frac{1}{N}\left\{\sum_{i=1}^{G}\sum_{j=1}^{N}\left(\frac{y_{i+}}{y_{++}}\right)^q\right\}^{1/(1-q)} = \frac{1}{N}\left\{\sum_{i=1}^{G}\sum_{j=1}^{N}(w_j p_{ij})^q\right\}^{1/(1-q)} \quad (q \neq 1) \quad (10)$$

$$^1D_\alpha = \lim_{q \to 1}{}^qD_\gamma = \exp\left\{-\sum_{i=1}^{G}\sum\left(\frac{y_{ij}}{y_{++}}\right)\log\left(\frac{y_{ij}}{y_{++}}\right) - \log(N)\right\} = \exp\left\{-\sum_{i=1}^{G}\sum_{j=1}^{N}(w_j p_{ij})\log(w_i p_{ij}) - \log(N)\right\} \quad (q = 1) \quad (11)$$

The computation of SNP beta-diversity can then be accomplished with Eqns. (7-11), *i.e.*, eqns. (7-8) for gamma diversity, (9) for beta-diversity and (10-11) for alpha-diversity.

Finally, we define a series of the Hill numbers for SNP diversity at different diversity order $q$=0, 1, 2,… as *SNP diversity profile*, whether it is alpha-, beta-, or gamma-diversity. The SNP diversity profile offers more comprehensive characterizations than the existing single diversity measure such as Shannon index or Simpson index because it captures the full spectrum of SNP diversity at different nonlinearity levels of diversity orders ($q$), corresponding to different levels of weights with SNP frequency distributions.

## The definitions for SNP similarities

To take advantages of the Hill numbers as SNP diversity measures, we also define Hill-numbers-based similarity measures for comparing multiple individuals in a population, multiple populations of a species, or multiple species in a community. We adopted the exactly same mathematical formulae originally used for defining community similarity measures in community ecology, as summarized in Chao *et al*. (2012, 2014) and Chiu *et al*. (2014). Chiu *et al*. (2014) found that the four existing similarity measures, Jaccard, Sørensen, Horn, Morisita-Horn in community ecology are actually some incarnations of the beta diversity ($^qD_\beta$) in the Hill numbers at different diversity order numbers, similar to the Hill numbers at different diversity order. Therefore, the four indexes form a series of *SNP similarity profile*, which may offer a more comprehensive and accurate characterization of the difference among individuals, among populations, or among species, depending on the basic genetic entity to be compared.

Here we briefly introduce the four similarity measures in the context of *N*-population of individuals. An advantage of using these similarity measures, rather than the beta-diversity directly is that they are 'normalized' to the range of [0, 1] by the number of individuals (*N*) (while beta-diversity of *N*-population ranges from 1 to *N*), and therefore they can be used to compare two groups with different numbers of individuals.

### (*i*) Local SNP Overlap ($C_{qN}$)

The *local SNP overlap* measure ($C_{qN}$) quantifies the effective average proportion of SNPs that are shared across all *N* individuals:

$$C_{qN} = \frac{\left(1/{}^qD_\beta\right)^{q-1} - (1/N)^{q-1}}{1 - (1/N)^{q-1}} \quad (12)$$

where $^qD_\beta$ is the SNP beta-diversity at order $q$ computed with Eqn. (8), $N$ is the number of individuals in the population. When $q$=0, $C_{qN}$ is actually the Sørensen similarity index; $q$=1, $C_{qN}$ is the Horn similarity index; $q$=2, $C_{qN}$ is the Morisita-Horn similarity index.

### (*ii*) Regional SNP overlap ($U_{qN}$)

The *regional SNP overlap* measure ($U_{qN}$) quantifies the effective proportion of shared SNPs in the pooled *N*-population:

$$U_{qN} = \frac{\left(1/{}^qD_\beta\right)^{1-q} - \left(1/N\right)^{1-q}}{1 - \left(1/N\right)^{1-q}} \quad (13)$$

When $q=0$, this statistic is equivalent to Jaccard similarity measure; $q=1$, it is equivalent to Horn similarity; $q=2$, it is equivalent to Morisita-Horn similarity index.

**(iii) SNP homogeneity measures ($S_{qN}$)**
$S_{qN}$ quantifies the *SNP homogeneity* (evenness) in an N-population:

$$S_{qN} = \frac{1/{}^qD_\beta - 1/N}{1 - 1/N} \quad (14)$$

When $q=0$, this statistic is equivalent to Jaccard similarity measure; $q=2$, it is equivalent to Morisita-Horn similarity index.

**(iv) SNP turnover complement ($V_{qN}$)**
The complement of $V_{qN}$ linearly quantifies the relative SNP turnover rate per individual. It represents the proportion of a typical individual that changes from one individual to another individual.

$$V_{qN} = \frac{N - {}^qD_\beta}{N - 1} = 1 - \frac{{}^qD_\beta - 1}{N - 1} \quad (15)$$

When $q=0$, this statistic is equivalent to Sørensen similarity measure; $q=2$, it is equivalent to Morisita-Horn similarity index.

# Computational Demonstration

## The datasets for the demonstration

We used the SNP datasets obtained through the whole genome sequencing of 9 individuals. Through a series of bioinformatics analyses, the list of all loci with SNP mutations, and the number of loci with SNP mutations on each chromosome were obtained from the raw sequence reads. The 9-cohort consists of 4 males and 5 females, including four individuals with lung cancer and 5 healthy relatives of the 4 cancer patients. Brief summary information on the 9-cohort is exhibited in Table S1, and detailed information on sequencing and bioinformatics procedures for obtaining the SNP datasets from the whole-genome sequencing of the DNA samples is referred to Kanwal *et al.* (2017).

## Demonstrations of the SNP Alpha-Diversity

Table S2 listed the SNP alpha-diversity for each chromosome of each individual in the 9-individual cohort. Table 1 below summarized the average SNP diversity of each chromosome per individual (averaged across 9 individuals). Besides illustrating the feasibility of SNP alpha-diversity, Table 1 shows that different chromosomes have different SNP patterns, as further exhibited in Fig 2. Therefore, the *SNP diversity profile* (the Hill numbers at different diversity orders) offers an effective tool to assess different mutation profiles on different chromosomes (as illustrated in this article), different individuals in a population, or different populations of a species.

Fig 2, which was plotted with the average SNP alpha-diversity (*y*-axis), averaged across 9 individuals, for each chromosome (*x*-axis), for each diversity order (one curve for each diversity

order *q*), respectively, shows the *diversity profiles* of each chromosome, and also the SNP diversity distribution pattern across chromosomes (*i.e.*, different chromosomes have different diversity profiles). In fact, one may similarly build SNP diversity profiles for each individual in the 9-individual cohort. What are displayed in Fig 2 are the averaged diversity profiles across 9 individuals in the cohort.

In Table 2, we further classify the SNP diversity levels of the human chromosomes as three groups: the high, medium and low, according to their alpha-diversities listed in Table 1. Since high diversity implies high variations, and the results in Table 2 should also reflect the variability levels of the human chromosomes in terms of the SNP mutations. For example, Table 2 shows that chromosome-19 has the second highest SNP alpha-diversity, only next to chromosome-1, which is also obvious in Fig 2. According to Grimwood *et al* (2004), chromosome-19 possesses the highest gene density of all human chromosomes, more than double the genome-wide average. In addition, chromosome-19 is the 3$^{rd}$ shortest (only next to chromsome-21 and chromosome-*y*) chromosome (https://www.ncbi.nlm.nih.gov/grc/human/data). It may be the exceptionally high gene density on a rather short chromosome that leads to the 2$^{nd}$ highest SNP diversity on chromosome-19.

**Table 1**. The mean (*per individual*) SNP alpha-diversity on each chromosome, averaged across the 9 individuals in the 9-cohort, summarized from Table S2

| Chromosome | | *Hill Numbers* | | | | |
|---|---|---|---|---|---|---|
| | | *q*=0 | *q*=1 | *q*=2 | *q*=3 | *q*=4 |
| Chr1 | Mean | 2198 | 774.22 | 388.17 | 250.54 | 189.67 |
| | Std. Err. | 4 | 2.91 | 3.27 | 2.98 | 2.66 |
| Chr2 | Mean | 1410 | 504.56 | 237.16 | 138.57 | 98.16 |
| | Std. Err. | 2 | 1.42 | 1.15 | 1.54 | 1.58 |
| Chr3 | Mean | 1245 | 426.63 | 218.33 | 144.27 | 110.66 |
| | Std. Err. | 2 | 3.06 | 2.66 | 2.51 | 2.49 |
| Chr4 | Mean | 856 | 332.52 | 182.06 | 125.09 | 99.05 |
| | Std. Err. | 1 | 2.37 | 2.01 | 1.75 | 1.61 |
| Chr5 | Mean | 962 | 351.58 | 198.78 | 139.68 | 110.93 |
| | Std. Err. | 4 | 2.03 | 1.99 | 2.22 | 2.38 |
| Chr6 | Mean | 1140 | 346.19 | 134.30 | 70.50 | 49.42 |
| | Std. Err. | 5 | 2.50 | 1.06 | 0.26 | 0.19 |
| Chr7 | Mean | 1063 | 319.05 | 159.61 | 107.31 | 83.86 |
| | Std. Err. | 4 | 1.86 | 1.93 | 1.82 | 1.70 |
| Chr8 | Mean | 763 | 208.97 | 64.91 | 31.77 | 22.52 |
| | Std. Err. | 2 | 1.72 | 1.14 | 0.66 | 0.46 |
| Chr9 | Mean | 859 | 269.08 | 100.44 | 50.95 | 35.55 |
| | Std. Err. | 2 | 1.48 | 1.55 | 1.14 | 0.84 |
| Chr10 | Mean | 904 | 298.22 | 149.07 | 99.11 | 77.20 |
| | Std. Err. | 2 | 2.15 | 2.39 | 2.49 | 2.51 |
| Chr11 | Mean | 1344 | 345.57 | 141.62 | 89.22 | 69.13 |
| | Std. Err. | 3 | 2.63 | 1.92 | 1.61 | 1.48 |
| Chr12 | Mean | 1127 | 406.30 | 217.66 | 145.87 | 112.17 |
| | Std. Err. | 2 | 1.94 | 1.90 | 1.71 | 1.49 |
| Chr13 | Mean | 448 | 162.72 | 92.43 | 66.46 | 54.28 |
| | Std. Err. | 1 | 0.98 | 1.04 | 1.10 | 1.13 |
| Chr14 | Mean | 658 | 227.85 | 111.63 | 70.40 | 53.17 |
| | Std. Err. | 1 | 1.54 | 2.14 | 2.14 | 2.00 |
| Chr15 | Mean | 707 | 258.39 | 142.91 | 100.36 | 80.65 |
| | Std. Err. | 3 | 1.52 | 1.44 | 1.38 | 1.41 |
| Chr16 | Mean | 854 | 178.10 | 49.80 | 27.91 | 21.28 |
| | Std. Err. | 4 | 0.95 | 0.51 | 0.34 | 0.29 |
| Chr17 | Mean | 1248 | 398.62 | 177.81 | 109.19 | 80.99 |
| | Std. Err. | 3 | 3.31 | 2.13 | 1.31 | 1.00 |
| Chr18 | Mean | 323 | 141.63 | 83.36 | 59.50 | 47.86 |
|  | Std. Err. | 1 | 1.20 | 1.49 | 1.75 | 1.85 |
| Chr19 | Mean | 1452 | 676.51 | 362.91 | 223.33 | 161.57 |
|  | Std. Err. | 3 | 4.31 | 6.13 | 7.63 | 7.79 |
| Chr20 | Mean | 606 | 182.60 | 65.86 | 36.87 | 27.24 |
|  | Std. Err. | 2 | 2.09 | 1.37 | 0.83 | 0.62 |
| Chr21 | Mean | 279 | 94.24 | 50.54 | 35.66 | 29.08 |
|  | Std. Err. | 1 | 0.64 | 0.80 | 0.81 | 0.79 |
| Chr22 | Mean | 518 | 214.21 | 113.32 | 75.98 | 59.88 |
|  | Std. Err. | 1 | 1.63 | 1.58 | 1.43 | 1.30 |
| ChrX | Mean | 725 | 194.54 | 71.80 | 41.25 | 30.71 |
|  | Std. Err. | 15 | 3.10 | 2.98 | 2.34 | 1.88 |
| ChrY | Mean | 34 | 16.185 | 9.198 | 6.855 | 5.860 |
|  | Std. Err. | 1 | 0.110 | 0.231 | 0.243 | 0.224 |
| Mean | Mean | 904 | 305.006 | 146.635 | 93.478 | 71.180 |
|  | Std. Error | 30 | 11.556 | 6.205 | 4.036 | 3.040 |

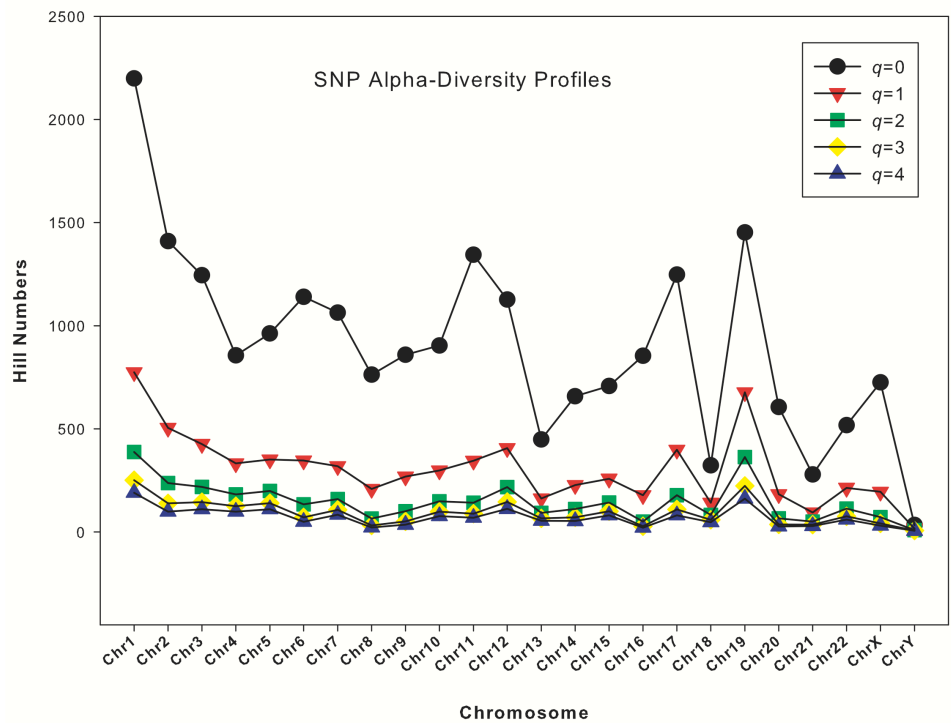

**Fig 2**. The SNP alpha-diversity profiles for each chromosome, averaged across 9-individuals

**Table 2**. Classifying the human chromosomes as the high, medium and low SNP-diversity groups, based on their alpha-diversities listed in Table 1

| Diversity Level | q=0 | | q=1 | | q=2 | | q=3 | | q=4 | |
|---|---|---|---|---|---|---|---|---|---|---|
|  | Chr. No | $^0D$ | Chr. No | $^1D$ | Chr. No | $^2D$ | Chr. No | $^3D$ | Chr. No | $^4D$ |
| **High SNP Diversity** | Chr1 | 2198 | Chr1 | 774.22 | Chr1 | 388.17 | Chr1 | 250.54 | Chr1 | 189.67 |
|  | Chr19 | 1452 | Chr19 | 676.51 | Chr19 | 362.91 | Chr19 | 223.33 | Chr19 | 161.57 |
|  | Chr2 | 1410 | Chr2 | 504.56 | Chr2 | 237.16 | Chr12 | 145.87 | Chr12 | 112.17 |
|  | Chr11 | 1344 | Chr3 | 426.63 | Chr3 | 218.33 | Chr3 | 144.27 | Chr5 | 110.93 |
|  | Chr17 | 1248 | Chr12 | 406.3 | Chr12 | 217.66 | Chr5 | 139.68 | Chr3 | 110.66 |
|  | Chr3 | 1245 | Chr17 | 398.62 | Chr5 | 198.78 | Chr2 | 138.57 | Chr4 | 99.05 |
|  | Chr6 | 1140 | Chr5 | 351.58 | Chr4 | 182.06 | Chr4 | 125.09 | Chr2 | 98.16 |
|  | Chr12 | 1127 | Chr6 | 346.19 | Chr17 | 177.81 | Chr17 | 109.19 | Chr7 | 83.86 |
| **Medium SNP Diversity** | Chr7 | 1063 | Chr11 | 345.57 | Chr7 | 159.61 | Chr7 | 107.31 | Chr17 | 80.99 |
|  | Chr5 | 962 | Chr4 | 332.52 | Chr10 | 149.07 | Chr15 | 100.36 | Chr15 | 80.65 |
|  | Chr10 | 904 | Chr7 | 319.05 | Chr15 | 142.91 | Chr10 | 99.11 | Chr10 | 77.20 |
|  | Chr9 | 859 | Chr10 | 298.22 | Chr11 | 141.62 | Chr11 | 89.22 | Chr11 | 69.13 |

| | | Chr4 | 856 | Chr9 | 269.08 | Chr6 | 134.30 | Chr22 | 75.98 | Chr22 | 59.88 |
| --- | --- | --- | --- | --- | --- | --- | --- | --- | --- | --- | --- |
| | | Chr16 | 854 | Chr15 | 258.39 | Chr22 | 113.32 | Chr6 | 70.50 | Chr13 | 54.28 |
| | | Chr8 | 763 | Chr14 | 227.85 | Chr14 | 111.63 | Chr14 | 70.40 | Chr14 | 53.17 |
| | | ChrX | 725 | Chr22 | 214.21 | Chr9 | 100.44 | Chr13 | 66.46 | Chr6 | 49.42 |
| | | Chr15 | 707 | Chr8 | 208.97 | Chr13 | 92.43 | Chr18 | 59.50 | Chr18 | 47.86 |
| | | Chr14 | 658 | ChrX | 194.54 | Chr18 | 83.36 | Chr9 | 50.95 | Chr9 | 35.55 |
| | | Chr20 | 606 | Chr20 | 182.6 | ChrX | 71.80 | ChrX | 41.25 | ChrX | 30.71 |
| Low | | Chr22 | 518 | Chr16 | 178.1 | Chr20 | 65.86 | Chr20 | 36.87 | Chr21 | 29.08 |
| SNP Diversity | | Chr13 | 448 | Chr13 | 162.72 | Chr8 | 64.91 | Chr21 | 35.66 | Chr20 | 27.24 |
| | | Chr18 | 323 | Chr18 | 141.63 | Chr21 | 50.54 | Chr8 | 31.77 | Chr8 | 22.52 |
| | | Chr21 | 279 | Chr21 | 94.24 | Chr16 | 49.80 | Chr16 | 27.91 | Chr16 | 21.28 |
| | | ChrY | 34 | ChrY | 16.185 | ChrY | 9.198 | ChrY | 6.855 | ChrY | 5.86 |

## Demonstrations of the SNP Beta-Diversity and Similarities

We demonstrate the computation of *SNP beta-diversity* with a slightly different scheme from the computation of SNP alpha-diversity. That is, we compute the pair-wise SNP beta-diversity and similarity for the same (numbered) chromosome between any two individuals in the 9-individual cohort. There are a total of 36 possible pairs among the 9 individuals. We compute the averages of the SNP beta-diversity or similarity values across the 36 pairs, and report the mean beta-diversity and similarity in Table 3 below.

Besides illustrating the feasibility and utility of proposed definitions for SNP beta-diversity and similarity measures, Table 3 shows an interesting phenomenon, which is expected from the nature of the *y*-chromosome, that is, the beta-diversity of *y*-chromosome at diversity order $q=0$ (*i.e.*, *SNP richness*) is generally higher than the beta-diversity of other chromosomes (including the *x*-chromosome) between two individuals.

The higher SNP beta-diversity of *y*-chromosome can be explained by the high mutation rate reported in the literature (Graves 2006, Lindblad-Toh *et al*. 2005). It was suggested that the two environmental factors that y-chromosome is housed are responsible for its high mutation rate (https://en.wikipedia.org/wiki/Y_chromosome). One is that *y*-chromosome is passed exclusively through sperm, which undergoes multiple cell divisions during gametogenesis. The cellular divisions greatly increase the probability to accumulate base pair mutations, *i.e.,* the frequency of SNPs. Second, the testis where sperms are stored is a highly oxidative environment that facilitates further mutation. These two environmental factors combined put *y*-chromosome at a higher probability of SNP occurrence than the rest of the human genome. However, it should be noted that the high SNP beta-diversity of *y*-chromosome only occurred at diversity order $q=0$, *i.e.,* the *SNP richness* or the number of loci with SNP. Since the focus of this study is to demonstrate the feasibility of our proposal for measuring SNP diversity and similarity, we stop further pursuing this interesting finding, but suggest to further confirm our finding with more extensive SNP datasets in future.

**Table 3**. The means of pair-wise SNP beta-diversity and similarity measures between any two individuals in the 9-subject cohort, for each chromosome

| Chromosome | $q=0$ | | | | | $q=1$ | | | | | $q=2$ | | | | |
| --- | --- | --- | --- | --- | --- | --- | --- | --- | --- | --- | --- | --- | --- | --- | --- |
| | *Beta* | Four Similarity Measures | | | | *Beta* | Four Similarity Measures | | | | *Beta* | Four Similarity Measures | | | |
| | | $C_q$ | $U_q$ | $S_q$ | $V_q$ | | $C_q$ | $U_q$ | $S_q$ | $V_q$ | | $C_q$ | $U_q$ | $S_q$ | $V_q$ |
| Chr1 | 1.037 | 0.963 | 0.929 | 0.929 | 0.963 | 1.023 | 0.967 | 0.967 | 0.955 | 0.977 | 1.015 | 0.970 | 0.985 | 0.970 | 0.985 |
| Chr2 | 1.028 | 0.972 | 0.945 | 0.945 | 0.972 | 1.023 | 0.967 | 0.967 | 0.954 | 0.977 | 1.014 | 0.972 | 0.986 | 0.972 | 0.986 |
| Chr3 | 1.034 | 0.966 | 0.935 | 0.935 | 0.966 | 1.021 | 0.970 | 0.970 | 0.959 | 0.979 | 1.014 | 0.973 | 0.986 | 0.973 | 0.986 |
| Chr4 | 1.021 | 0.979 | 0.959 | 0.959 | 0.979 | 1.020 | 0.971 | 0.971 | 0.960 | 0.980 | 1.016 | 0.969 | 0.984 | 0.969 | 0.984 |
| Chr5 | 1.042 | 0.958 | 0.920 | 0.920 | 0.958 | 1.030 | 0.958 | 0.958 | 0.943 | 0.970 | 1.020 | 0.962 | 0.980 | 0.962 | 0.980 |
| Chr6 | 1.034 | 0.966 | 0.933 | 0.933 | 0.966 | 1.023 | 0.968 | 0.968 | 0.956 | 0.977 | 1.010 | 0.980 | 0.990 | 0.980 | 0.990 |
| Chr 7 | 1.042 | 0.958 | 0.920 | 0.920 | 0.958 | 1.023 | 0.968 | 0.968 | 0.956 | 0.977 | 1.017 | 0.967 | 0.983 | 0.967 | 0.983 |
| Chr8 | 1.033 | 0.967 | 0.937 | 0.937 | 0.967 | 1.022 | 0.969 | 0.969 | 0.957 | 0.978 | 1.008 | 0.984 | 0.992 | 0.984 | 0.992 |
| Chr9 | 1.042 | 0.958 | 0.919 | 0.919 | 0.958 | 1.023 | 0.967 | 0.967 | 0.955 | 0.977 | 1.011 | 0.979 | 0.989 | 0.979 | 0.989 |
| Chr10 | 1.031 | 0.969 | 0.939 | 0.939 | 0.969 | 1.023 | 0.967 | 0.967 | 0.955 | 0.977 | 1.014 | 0.973 | 0.986 | 0.973 | 0.986 |
| Chr11 | 1.038 | 0.962 | 0.926 | 0.926 | 0.962 | 1.022 | 0.968 | 0.968 | 0.956 | 0.978 | 1.012 | 0.976 | 0.988 | 0.976 | 0.988 |
| Chr12 | 1.030 | 0.970 | 0.941 | 0.941 | 0.970 | 1.023 | 0.968 | 0.968 | 0.956 | 0.977 | 1.016 | 0.968 | 0.984 | 0.968 | 0.984 |
| Chr13 | 1.025 | 0.975 | 0.952 | 0.952 | 0.975 | 1.015 | 0.978 | 0.978 | 0.970 | 0.985 | 1.011 | 0.978 | 0.989 | 0.978 | 0.989 |
| Chr14 | 1.083 | 0.917 | 0.857 | 0.857 | 0.917 | 1.052 | 0.929 | 0.929 | 0.906 | 0.948 | 1.027 | 0.948 | 0.973 | 0.948 | 0.973 |
| Chr15 | 1.038 | 0.962 | 0.927 | 0.927 | 0.962 | 1.023 | 0.967 | 0.967 | 0.955 | 0.977 | 1.016 | 0.969 | 0.984 | 0.969 | 0.984 |
| Chr16 | 1.042 | 0.958 | 0.919 | 0.919 | 0.958 | 1.024 | 0.966 | 0.966 | 0.953 | 0.976 | 1.005 | 0.989 | 0.995 | 0.989 | 0.995 |
| Chr17 | 1.043 | 0.957 | 0.917 | 0.917 | 0.957 | 1.029 | 0.959 | 0.959 | 0.944 | 0.971 | 1.014 | 0.973 | 0.986 | 0.973 | 0.986 |
| Chr18 | 1.019 | 0.981 | 0.962 | 0.962 | 0.981 | 1.017 | 0.975 | 0.975 | 0.966 | 0.983 | 1.016 | 0.969 | 0.984 | 0.969 | 0.984 |
| Chr19 | 1.047 | 0.953 | 0.909 | 0.909 | 0.953 | 1.050 | 0.929 | 0.929 | 0.904 | 0.950 | 1.041 | 0.921 | 0.959 | 0.921 | 0.959 |
| Chr20 | 1.042 | 0.958 | 0.920 | 0.920 | 0.958 | 1.029 | 0.959 | 0.959 | 0.945 | 0.971 | 1.012 | 0.976 | 0.988 | 0.976 | 0.988 |
| Chr21 | 1.037 | 0.963 | 0.929 | 0.929 | 0.963 | 1.019 | 0.973 | 0.973 | 0.963 | 0.981 | 1.013 | 0.974 | 0.987 | 0.974 | 0.987 |
| Chr22 | 1.034 | 0.966 | 0.935 | 0.935 | 0.966 | 1.032 | 0.954 | 0.954 | 0.938 | 0.968 | 1.025 | 0.951 | 0.975 | 0.951 | 0.975 |
| ChrX | 1.087 | 0.913 | 0.841 | 0.841 | 0.913 | 1.058 | 0.919 | 0.919 | 0.891 | 0.942 | 1.064 | 0.883 | 0.936 | 0.883 | 0.936 |
| ChrY | 1.147 | 0.853 | 0.745 | 0.745 | 0.853 | 1.039 | 0.945 | 0.945 | 0.925 | 0.961 | 1.011 | 0.977 | 0.989 | 0.977 | 0.989 |
| Mean | 1.044 | 0.956 | 0.917 | 0.917 | 0.956 | 1.028 | 0.961 | 0.961 | 0.947 | 0.972 | 1.018 | 0.966 | 0.982 | 0.966 | 0.982 |
| Std. Error | 0.006 | 0.006 | 0.009 | 0.009 | 0.006 | 0.002 | 0.003 | 0.003 | 0.004 | 0.002 | 0.002 | 0.005 | 0.002 | 0.005 | 0.002 |

## Discussion

Nachman (2001) identified three fields where the study of SNPs is of critical importance: (*i*) SNPs hold great potential as markers for mapping polygenic disease loci, because the frequency and underlying patterns of the association among SNPs unrelated to disease is critical for interpreting patterns of linkage disequilibrium between markers and candidate disease genes. (*ii*) SNPs may shed light on human history, including relationships among ethnic groups, migrations and changes in population size. (*iii*) The SNP distribution may teach us about the relative importance of forces such as selection, mutation, migration, recombination and genetic drift, hence, can help us understand the nature of the evolutionary process at the molecular level. The SNP diversity and similarity profiles (measures) we introduced previously can, directly or indirectly, facilitate the exploration of the above three fields, and therefore find important practical applications in population genetics and genomics.

SNPs may occur in coding sequences of genes, non-coding regions of genes, or in the inter-genic regions. Accordingly, the SNP diversity defined in this article can be applied separately to the three types of SNP occurrence regions. For demonstrative purpose, we did not distinguish the three types in this article, but all the definitions and computational procedures (including the R-program) presented in previous sections can be directly applied to separate measuring of the SNP diversities. The only, but minor, difference would be in the data preparation step, *i.e.*, the preparatory calculation of $p_i$ according to the region chosen, either coding, non-coding, inter-genic, or the whole locus.

We demonstrated SNP alpha-diversity with single chromosome as the basic *genetic entity*, which could also just be the genome of an individual. That is, the SNP diversity for the genome of an individual can be computed similarly with our definitions. Again for demonstrative purpose, we computed the *pair-wise* SNP beta-diversity for the *same-numbered* two chromosomes from two different individuals. Of course, the pair-wise SNP beta-diversity for the genomes of two individuals can be computed if the genetic entity chosen is the whole genome. Similarly, SNP beta-diversity may be computed for multiple (*N*) individuals, as defined previously.

Besides defining and demonstrating SNP diversities, we also defined four similarity measures, all of which are based on the Hill numbers and are some functions of the multiplicative beta-diversity. An advantage of using the similarity measures, rather than beta-diversity directly, is that the similarity measures are normalized to [0, 1] interval, hence, independent of the number of individuals in the population.

The *y*-chromosome exhibited higher pair-wise beta-diversity and lower similarity at the diversity order $q=0$, *i.e.*, SNP *richness*. This can be explained with the biological reality that *y*-chromosome has higher mutation rate than other chromosomes due to the environment it resides (Graves 2006, Lindblad-Toh *et al*. 2005). This finding highlights a utility of our proposal introduced in this study.

As a side note, we also computed the average diversity per chromosome for each individual (Suppl. Table S3) for demonstrative purpose. The average is computed across all chromosomes of an individual, and a single value of the average (actually a series of the Hill numbers) is obtained for each individual. Although the single diversity may be utilized to compare different individuals in a population in terms of their SNP distribution pattern, it is less useful than directly computing the genome-scale SNP diversity for each individual, which we skipped in this article but can be easily performed with the computational procedures and program provided. We further tested the difference between the healthy and diseased groups or between the male and female groups (Suppl. Table S4). No significance was found in either of the tests (all *p-values*>0.05). However, the lack of significant difference may be due to our skip of distinguishing the three types of SNP regions (coding, non-coding and inter-genic). Another possible factor could be the small sample size of our demonstrative datasets. Since these additional computations and significance tests are hardly relevant to the objective of this study, both the computational procedures and program we employed should have sufficiently demonstrated the validity and utility of our proposal for measuring SNP diversity and similarity with Hill numbers.

Recently, Gaggiotti et al (2018) developed a unifying framework for measuring biodiversity from genes to ecosystems by standardizing on the Hill number at diversity order $q=1$, which is a transformation of Shannon diversity index. Their simplification is necessary to develop a more generalized framework, but it does not obsolete the novelty of our work here. This is because, at a specific level (the genome level of an individual), Hill numbers at difference orders ($q=0$, 1, 2,…) are still necessary to present a comprehensive diversity profile due to the complexity of the issues involved, as demonstrated in previous sections.

## Online Supplementary Materials (OSM)
**Table S1**. The datasets utilized for illustrating the SNP diversity and similarity
**Table S2**. The SNP alpha-diversity on each chromosome of each individual in the 9-cohort
**Table S3**. The mean (*per chromosome*) SNP alpha-diversity for each individual
**Table S4**. The Wilcoxon significance test for the differences between the healthy and diseased or between the male and female individuals

**Table S1**. The datasets utilized for illustrating the SNP diversity and similarity

| Family | ID | Sex | Age | Disease | Smoking | Postscript |
|---|---|---|---|---|---|---|
| Family 1 | BH1700 | F | 43 | Cancer | No | |
| | BH1696 | F | 21 | Healthy | No | Daughter of BH1700 |
| Family 2 | BH1701 | M | 58 | Cancer | Yes | |
| | BH1704 | M | 28 | Healthy | Unknown | Son of BH1701 |
| | BH1705 | M | 26 | Healthy | Unknown | Son of BH1701 |
| Family 3 | BH1706 | F | 34 | Cancer | No | |
| Family 4 | BH1708 | M | 66 | Cancer | Yes | |
| | BH1713 | F | 43 | Healthy | No | Niece of BH1708 |
| | BH1716 | F | 40 | Healthy | No | Niece of BH1708 |

**Table S2**. The SNP alpha-diversity on each chromosome of each individual from the 9-individual cohort

| Sample ID | $q=0$ | $q=1$ | $q=2$ | $q=3$ | $q=4$ |
|---|---|---|---|---|---|
| BH1696_chr1 | 2205 | 785.31 | 398.38 | 259.78 | 197.90 |
| BH1700_chr1 | 2201 | 781.73 | 400.79 | 263.01 | 200.40 |
| BH1701_chr1 | 2178 | 770.47 | 383.63 | 247.09 | 187.74 |
| BH1704_chr1 | 2183 | 763.84 | 375.55 | 237.20 | 176.29 |
| BH1705_chr1 | 2207 | 760.56 | 375.15 | 241.36 | 183.15 |
| BH1706_chr1 | 2205 | 774.42 | 386.30 | 248.09 | 186.69 |
| BH1708_chr1 | 2212 | 785.25 | 400.19 | 260.25 | 197.53 |
| BH1713_chr1 | 2195 | 772.95 | 388.35 | 252.32 | 192.52 |
| BH1716_chr1 | 2196 | 773.43 | 385.23 | 245.76 | 184.83 |
| Mean | 2198 | 774.22 | 388.17 | 250.54 | 189.67 |
| Std. Error | 4 | 2.91 | 3.27 | 2.98 | 2.66 |
| BH1696_chr2 | 1415 | 504.44 | 233.21 | 135.34 | 95.76 |
| BH1700_chr2 | 1411 | 505.46 | 235.94 | 138.98 | 99.10 |
| BH1701_chr2 | 1406 | 503.45 | 232.72 | 131.46 | 91.04 |
| BH1704_chr2 | 1408 | 506.98 | 237.04 | 133.71 | 91.96 |
| BH1705_chr2 | 1415 | 506.26 | 238.44 | 141.32 | 101.92 |
| BH1706_chr2 | 1395 | 512.08 | 238.96 | 137.53 | 96.34 |
| BH1708_chr2 | 1414 | 498.50 | 236.90 | 141.36 | 100.97 |
| BH1713_chr2 | 1414 | 498.23 | 236.69 | 140.82 | 101.14 |
| BH1716_chr2 | 1413 | 505.67 | 244.50 | 146.60 | 105.19 |
| Mean | 1410 | 504.56 | 237.16 | 138.57 | 98.16 |
| Std. Error | 2 | 1.42 | 1.15 | 1.54 | 1.58 |
| BH1696_chr3 | 1236 | 420.97 | 216.24 | 144.72 | 112.05 |
| BH1700_chr3 | 1233 | 413.26 | 212.68 | 143.55 | 112.04 |
| BH1701_chr3 | 1251 | 428.83 | 224.37 | 150.69 | 116.53 |
| BH1704_chr3 | 1250 | 421.78 | 215.67 | 142.10 | 108.28 |
| BH1705_chr3 | 1245 | 427.99 | 221.51 | 147.40 | 114.01 |
| BH1706_chr3 | 1245 | 447.15 | 234.78 | 155.93 | 118.95 |
| BH1708_chr3 | 1242 | 426.91 | 218.72 | 147.45 | 116.63 |
| BH1713_chr3 | 1255 | 423.92 | 206.81 | 130.86 | 97.47 |
| BH1716_chr3 | 1250 | 428.82 | 214.14 | 135.73 | 100.00 |
| Mean | 1245 | 426.63 | 218.33 | 144.27 | 110.66 |
| Std. Error | 2 | 3.06 | 2.66 | 2.51 | 2.49 |
| BH1696_chr4 | 857 | 342.91 | 191.96 | 132.84 | 105.37 |
| BH1700_chr4 | 847 | 330.88 | 178.15 | 118.83 | 91.80 |
| BH1701_chr4 | 857 | 327.09 | 175.11 | 119.49 | 94.97 |
| BH1704_chr4 | 856 | 333.31 | 184.97 | 129.27 | 104.08 |
| BH1705_chr4 | 856 | 324.76 | 177.13 | 122.31 | 97.52 |
| BH1706_chr4 | 857 | 325.31 | 179.65 | 126.28 | 101.94 |
| BH1708_chr4 | 857 | 330.85 | 182.88 | 126.67 | 100.00 |
| BH1713_chr4 | 857 | 332.92 | 178.06 | 119.51 | 93.48 |
| BH1716_chr4 | 857 | 344.66 | 190.65 | 130.57 | 102.32 |
| Mean | 856 | 332.52 | 182.06 | 125.09 | 99.05 |
| Std. Error | 1 | 2.37 | 2.01 | 1.75 | 1.61 |
| BH1696_chr5 | 955 | 349.06 | 198.27 | 140.45 | 112.08 |
| BH1700_chr5 | 960 | 355.94 | 208.49 | 152.13 | 124.33 |
| BH1701_chr5 | 946 | 340.90 | 188.56 | 129.10 | 100.47 |
| BH1704_chr5 | 984 | 357.62 | 195.97 | 133.13 | 102.30 |
| BH1705_chr5 | 971 | 350.60 | 199.36 | 140.51 | 111.82 |
| BH1706_chr5 | 957 | 356.19 | 201.00 | 139.90 | 110.06 |
| BH1708_chr5 | 966 | 357.74 | 203.51 | 142.53 | 113.22 |
| BH1713_chr5 | 966 | 343.80 | 192.31 | 135.44 | 107.98 |
| BH1716_chr5 | 954 | 352.39 | 201.55 | 143.94 | 116.13 |
| Mean | 962 | 351.58 | 198.78 | 139.68 | 110.93 |
| Std. Error | 4 | 2.03 | 1.99 | 2.22 | 2.38 |
| BH1696_chr6 | 1153 | 342.57 | 130.71 | 68.88 | 48.57 |
| BH1700_chr6 | 1160 | 354.18 | 137.05 | 70.85 | 48.97 |
| BH1701_chr6 | 1135 | 346.82 | 135.31 | 71.07 | 49.99 |
| BH1704_chr6 | 1105 | 337.24 | 132.80 | 70.29 | 49.45 |
| BH1705_chr6 | 1129 | 342.62 | 132.67 | 70.26 | 49.58 |
| BH1706_chr6 | 1137 | 359.57 | 140.36 | 71.53 | 48.93 |
| BH1708_chr6 | 1148 | 348.62 | 135.76 | 70.63 | 49.18 |
| BH1713_chr6 | 1146 | 336.61 | 130.39 | 69.94 | 49.86 |
| BH1716_chr6 | 1148 | 347.46 | 133.68 | 71.04 | 50.28 |
| Mean | 1140 | 346.19 | 134.30 | 70.50 | 49.42 |
| Std. Error | 5 | 2.50 | 1.06 | 0.26 | 0.19 |
| BH1696_chr7 | 1075 | 320.42 | 159.10 | 105.66 | 81.28 |
| BH1700_chr7 | 1071 | 316.32 | 158.25 | 105.65 | 81.80 |
| BH1701_chr7 | 1042 | 316.71 | 158.53 | 106.47 | 82.95 |
| BH1704_chr7 | 1053 | 316.66 | 156.09 | 104.77 | 82.50 |
| BH1705_chr7 | 1056 | 318.43 | 162.49 | 111.53 | 88.39 |
| BH1706_chr7 | 1065 | 325.97 | 165.88 | 113.24 | 89.78 |
| BH1708_chr7 | 1076 | 328.48 | 168.92 | 115.57 | 91.31 |
| BH1713_chr7 | 1069 | 309.52 | 148.75 | 97.25 | 74.93 |
| BH1716_chr7 | 1060 | 318.91 | 158.50 | 105.65 | 81.79 |
| Mean | 1063 | 319.05 | 159.61 | 107.31 | 83.86 |
| Std. Error | 4 | 1.86 | 1.93 | 1.82 | 1.70 |
| BH1696_chr8 | 757 | 217.27 | 66.85 | 31.76 | 22.33 |
| BH1700_chr8 | 767 | 217.49 | 72.04 | 35.52 | 25.01 |
| BH1701_chr8 | 765 | 206.21 | 65.60 | 32.78 | 23.30 |
| BH1704_chr8 | 771 | 204.66 | 64.41 | 31.97 | 22.70 |
| BH1705_chr8 | 773 | 204.91 | 62.65 | 30.85 | 21.93 |
| BH1706_chr8 | 766 | 210.10 | 66.56 | 33.35 | 23.77 |
| BH1708_chr8 | 758 | 204.88 | 62.86 | 30.68 | 21.82 |
| BH1713_chr8 | 759 | 205.80 | 60.25 | 28.79 | 20.42 |
| BH1716_chr8 | 751 | 209.41 | 62.96 | 30.23 | 21.39 |
| Mean | 763 | 208.97 | 64.91 | 31.77 | 22.52 |
| Std. Error | 2 | 1.72 | 1.14 | 0.66 | 0.46 |
| BH1696_chr9 | 869 | 259.83 | 91.92 | 45.20 | 31.27 |
| BH1700_chr9 | 856 | 269.62 | 103.50 | 52.51 | 36.31 |
| BH1701_chr9 | 853 | 268.09 | 101.55 | 51.97 | 36.43 |
| BH1704_chr9 | 858 | 273.15 | 99.73 | 49.26 | 34.10 |
| BH1705_chr9 | 857 | 274.86 | 99.03 | 48.80 | 33.82 |
| BH1706_chr9 | 858 | 265.14 | 96.42 | 48.72 | 34.22 |
| BH1708_chr9 | 857 | 270.19 | 99.19 | 50.76 | 35.81 |
| BH1713_chr9 | 861 | 271.12 | 105.87 | 55.10 | 38.59 |
| BH1716_chr9 | 862 | 269.67 | 106.74 | 56.20 | 39.44 |
| Mean | 859 | 269.08 | 100.44 | 50.95 | 35.55 |
| Std. Error | 2 | 1.48 | 1.55 | 1.14 | 0.84 |
| BH1696_chr10 | 904 | 297.79 | 150.73 | 101.48 | 79.87 |
| BH1700_chr10 | 907 | 300.53 | 151.11 | 99.99 | 77.34 |
| BH1701_chr10 | 908 | 306.75 | 159.07 | 109.30 | 87.18 |
| BH1704_chr10 | 906 | 303.23 | 153.04 | 103.09 | 80.97 |
| BH1705_chr10 | 912 | 301.84 | 155.24 | 107.53 | 86.92 |
| BH1706_chr10 | 900 | 300.50 | 151.54 | 100.17 | 77.27 |
| BH1708_chr10 | 901 | 297.45 | 143.51 | 92.07 | 69.94 |
| BH1713_chr10 | 897 | 287.45 | 138.57 | 88.40 | 66.61 |
| BH1716_chr10 | 905 | 288.47 | 138.84 | 89.92 | 68.69 |
| Mean | 904 | 298.22 | 149.07 | 99.11 | 77.20 |
| Std. Error | 2 | 2.15 | 2.39 | 2.49 | 2.51 |
| BH1696_chr11 | 1331 | 338.87 | 133.84 | 81.35 | 61.56 |
| BH1700_chr11 | 1342 | 336.59 | 134.76 | 83.74 | 64.01 |
| BH1701_chr11 | 1339 | 347.27 | 145.23 | 91.61 | 70.50 |
| BH1704_chr11 | 1341 | 341.85 | 139.81 | 88.25 | 68.80 |
| BH1705_chr11 | 1341 | 348.83 | 146.50 | 93.70 | 73.26 |
| BH1706_chr11 | 1364 | 359.93 | 148.94 | 94.19 | 73.09 |
| BH1708_chr11 | 1334 | 339.39 | 137.11 | 85.57 | 65.98 |
| BH1713_chr11 | 1347 | 342.29 | 140.31 | 89.90 | 70.78 |
| BH1716_chr11 | 1353 | 355.11 | 148.06 | 94.63 | 74.21 |
| Mean | 1344 | 345.57 | 141.62 | 89.22 | 69.13 |
| Std. Error | 3 | 2.63 | 1.92 | 1.61 | 1.48 |
| BH1696_chr12 | 1122 | 411.49 | 217.82 | 144.30 | 110.42 |
| BH1700_chr12 | 1118 | 410.97 | 222.93 | 150.93 | 116.78 |
| BH1701_chr12 | 1129 | 414.96 | 225.44 | 151.26 | 115.72 |
| BH1704_chr12 | 1133 | 402.26 | 208.79 | 136.68 | 103.98 |
| BH1705_chr12 | 1140 | 404.91 | 216.11 | 145.02 | 112.03 |
| BH1706_chr12 | 1123 | 400.72 | 214.45 | 142.81 | 108.66 |
| BH1708_chr12 | 1132 | 410.23 | 223.80 | 151.92 | 117.56 |

| Sample | | | | | |
|---|---|---|---|---|---|
| BH1713_chr12 | 1122 | 403.61 | 218.49 | 148.25 | 114.95 |
| BH1716_chr12 | 1123 | 397.57 | 211.12 | 141.65 | 109.41 |
| Mean | 1127 | 406.30 | 217.66 | 145.87 | 112.17 |
| Std. Error | 2 | 1.94 | 1.90 | 1.71 | 1.49 |
| BH1696_chr13 | 444 | 164.19 | 91.84 | 65.06 | 52.46 |
| BH1700_chr13 | 445 | 161.71 | 91.04 | 64.62 | 52.00 |
| BH1701_chr13 | 454 | 161.05 | 90.07 | 63.52 | 50.98 |
| BH1704_chr13 | 448 | 158.01 | 88.85 | 64.06 | 52.81 |
| BH1705_chr13 | 446 | 159.92 | 90.39 | 65.12 | 53.57 |
| BH1706_chr13 | 449 | 163.36 | 91.71 | 64.10 | 51.14 |
| BH1708_chr13 | 448 | 168.04 | 99.09 | 73.13 | 60.61 |
| BH1713_chr13 | 448 | 163.40 | 94.34 | 69.28 | 57.45 |
| BH1716_chr13 | 449 | 164.77 | 94.57 | 69.23 | 57.51 |
| Mean | 448 | 162.72 | 92.43 | 66.46 | 54.28 |
| Std. Error | 1 | 0.98 | 1.04 | 1.10 | 1.13 |
| BH1696_chr14 | 656 | 235.72 | 121.09 | 79.06 | 60.65 |
| BH1700_chr14 | 660 | 227.39 | 115.39 | 76.32 | 59.49 |
| BH1701_chr14 | 656 | 225.78 | 107.76 | 64.85 | 47.35 |
| BH1704_chr14 | 665 | 221.68 | 100.44 | 58.66 | 42.25 |
| BH1705_chr14 | 656 | 224.67 | 110.04 | 69.80 | 52.64 |
| BH1706_chr14 | 652 | 232.44 | 115.96 | 72.32 | 54.09 |
| BH1708_chr14 | 653 | 229.22 | 114.38 | 72.55 | 54.48 |
| BH1713_chr14 | 658 | 222.94 | 104.71 | 65.40 | 49.67 |
| BH1716_chr14 | 663 | 230.85 | 114.87 | 74.64 | 57.95 |
| Mean | 658 | 227.85 | 111.63 | 70.40 | 53.17 |
| Std. Error | 1 | 1.54 | 2.14 | 2.14 | 2.00 |
| BH1696_chr15 | 709 | 265.63 | 150.67 | 107.79 | 87.92 |
| BH1700_chr15 | 719 | 264.65 | 145.45 | 100.43 | 79.48 |
| BH1701_chr15 | 706 | 257.22 | 140.68 | 98.80 | 79.65 |
| BH1704_chr15 | 703 | 253.56 | 140.78 | 101.07 | 83.03 |
| BH1705_chr15 | 699 | 251.60 | 135.62 | 93.50 | 74.01 |
| BH1706_chr15 | 709 | 259.40 | 141.98 | 98.75 | 78.73 |
| BH1708_chr15 | 693 | 258.58 | 146.98 | 105.25 | 85.72 |
| BH1713_chr15 | 712 | 257.56 | 142.77 | 99.67 | 79.49 |
| BH1716_chr15 | 712 | 257.28 | 141.30 | 97.96 | 77.83 |
| Mean | 707 | 258.39 | 142.91 | 100.36 | 80.65 |
| Std. Error | 3 | 1.52 | 1.44 | 1.38 | 1.41 |
| BH1696_chr16 | 863 | 176.52 | 49.63 | 28.18 | 21.62 |
| BH1700_chr16 | 873 | 179.10 | 49.30 | 27.30 | 20.67 |
| BH1701_chr16 | 841 | 177.70 | 49.50 | 27.70 | 21.13 |
| BH1704_chr16 | 854 | 182.59 | 51.79 | 29.22 | 22.36 |
| BH1705_chr16 | 838 | 177.35 | 50.41 | 28.67 | 22.06 |
| BH1706_chr16 | 853 | 181.47 | 51.77 | 28.76 | 21.81 |
| BH1708_chr16 | 855 | 178.68 | 48.65 | 26.61 | 20.00 |
| BH1713_chr16 | 859 | 173.04 | 46.87 | 26.27 | 20.07 |
| BH1716_chr16 | 852 | 176.46 | 50.29 | 28.47 | 21.80 |
| Mean | 854 | 178.10 | 49.80 | 27.91 | 21.28 |
| Std. Error | 4 | 0.95 | 0.51 | 0.34 | 0.29 |
| BH1696_chr17 | 1255 | 386.01 | 167.81 | 103.79 | 77.93 |
| BH1700_chr17 | 1239 | 400.51 | 183.14 | 112.62 | 83.06 |
| BH1701_chr17 | 1245 | 390.78 | 171.89 | 103.94 | 76.13 |
| BH1704_chr17 | 1250 | 416.22 | 187.73 | 114.98 | 85.39 |
| BH1705_chr17 | 1252 | 401.73 | 179.26 | 111.10 | 83.49 |
| BH1706_chr17 | 1253 | 410.18 | 184.09 | 112.54 | 83.34 |
| BH1708_chr17 | 1251 | 390.34 | 173.30 | 107.05 | 79.39 |
| BH1713_chr17 | 1254 | 399.70 | 176.52 | 107.97 | 79.86 |
| BH1716_chr17 | 1233 | 392.07 | 176.59 | 108.76 | 80.30 |
| Mean | 1248 | 398.62 | 177.81 | 109.19 | 80.99 |
| Std. Error | 3 | 3.31 | 2.13 | 1.31 | 1.00 |
| BH1696_chr18 | 322 | 139.73 | 81.52 | 57.34 | 45.47 |
| BH1700_chr18 | 321 | 133.86 | 74.31 | 49.91 | 38.46 |
| BH1701_chr18 | 322 | 144.99 | 83.34 | 56.62 | 43.89 |
| BH1704_chr18 | 323 | 141.31 | 81.53 | 56.87 | 44.79 |
| BH1705_chr18 | 323 | 140.46 | 81.12 | 57.24 | 45.58 |
| BH1706_chr18 | 325 | 145.26 | 89.46 | 66.96 | 55.90 |
| BH1708_chr18 | 324 | 145.23 | 87.55 | 64.00 | 52.43 |
| BH1713_chr18 | 326 | 142.65 | 86.14 | 63.43 | 52.10 |
| BH1716_chr18 | 325 | 141.21 | 85.29 | 63.12 | 52.12 |
| Mean | 323 | 141.63 | 83.36 | 59.50 | 47.86 |
| Std. Error | 1 | 1.20 | 1.49 | 1.75 | 1.85 |
| BH1696_chr19 | 1450 | 674.00 | 349.87 | 202.90 | 139.35 |
| BH1700_chr19 | 1449 | 698.27 | 373.85 | 217.99 | 150.49 |
| BH1701_chr19 | 1460 | 677.20 | 354.39 | 209.76 | 147.23 |
| BH1704_chr19 | 1457 | 679.30 | 360.99 | 218.55 | 156.08 |
| BH1705_chr19 | 1447 | 659.31 | 344.81 | 210.06 | 151.52 |
| BH1706_chr19 | 1461 | 694.63 | 397.16 | 263.26 | 200.07 |
| BH1708_chr19 | 1463 | 670.55 | 345.72 | 205.12 | 145.73 |
| BH1713_chr19 | 1440 | 671.61 | 384.66 | 261.19 | 202.56 |
| BH1716_chr19 | 1442 | 663.71 | 354.75 | 221.11 | 161.12 |
| Mean | 1452 | 676.51 | 362.91 | 223.33 | 161.57 |
| Std. Error | 3 | 4.31 | 6.13 | 7.63 | 7.79 |
| BH1696_chr20 | 604 | 185.96 | 65.36 | 35.92 | 26.50 |
| BH1700_chr20 | 607 | 180.17 | 62.79 | 34.70 | 25.61 |
| BH1701_chr20 | 605 | 179.13 | 62.78 | 34.60 | 25.39 |
| BH1704_chr20 | 596 | 178.00 | 64.93 | 36.90 | 27.54 |
| BH1705_chr20 | 607 | 172.05 | 58.38 | 32.83 | 24.47 |
| BH1706_chr20 | 601 | 180.09 | 68.36 | 40.03 | 30.15 |
| BH1708_chr20 | 613 | 189.88 | 69.66 | 38.69 | 28.40 |
| BH1713_chr20 | 608 | 186.81 | 69.67 | 39.04 | 28.60 |
| BH1716_chr20 | 612 | 191.29 | 70.79 | 39.08 | 28.48 |
| Mean | 606 | 182.60 | 65.86 | 36.87 | 27.24 |
| Std. Error | 2 | 2.09 | 1.37 | 0.83 | 0.62 |
| BH1696_chr21 | 279 | 94.99 | 51.60 | 37.11 | 30.77 |
| BH1700_chr21 | 278 | 90.79 | 49.35 | 35.35 | 29.00 |
| BH1701_chr21 | 276 | 94.19 | 50.91 | 35.58 | 28.62 |
| BH1704_chr21 | 276 | 94.75 | 51.22 | 35.61 | 28.47 |
| BH1705_chr21 | 278 | 94.63 | 48.84 | 33.18 | 26.35 |
| BH1706_chr21 | 273 | 97.37 | 55.19 | 40.70 | 34.13 |
| BH1708_chr21 | 286 | 91.96 | 46.47 | 32.19 | 26.33 |
| BH1713_chr21 | 281 | 95.33 | 51.56 | 36.40 | 29.66 |
| BH1716_chr21 | 280 | 94.14 | 49.75 | 34.78 | 28.40 |
| Mean | 279 | 94.24 | 50.54 | 35.66 | 29.08 |
| Std. Error | 1 | 0.64 | 0.80 | 0.81 | 0.79 |
| BH1696_chr22 | 520 | 219.24 | 118.80 | 81.58 | 64.82 |
| BH1700_chr22 | 519 | 213.61 | 112.32 | 74.89 | 59.13 |
| BH1701_chr22 | 516 | 210.10 | 110.92 | 75.25 | 59.95 |
| BH1704_chr22 | 522 | 210.93 | 109.92 | 72.53 | 56.12 |
| BH1705_chr22 | 514 | 207.54 | 105.56 | 69.20 | 54.23 |
| BH1706_chr22 | 523 | 218.43 | 117.93 | 80.89 | 64.68 |
| BH1708_chr22 | 519 | 219.41 | 116.81 | 77.31 | 60.36 |
| BH1713_chr22 | 518 | 219.23 | 118.15 | 79.96 | 63.46 |
| BH1716_chr22 | 510 | 209.40 | 109.46 | 72.19 | 56.19 |
| Mean | 518 | 214.21 | 113.32 | 75.98 | 59.88 |
| Std. Error | 1 | 1.63 | 1.58 | 1.43 | 1.30 |
| BH1696_chrX | 763 | 200.88 | 72.49 | 40.19 | 29.19 |
| BH1700_chrX | 755 | 197.80 | 76.74 | 45.51 | 34.29 |
| BH1701_chrX | 684 | 185.39 | 71.48 | 41.35 | 30.57 |
| BH1704_chrX | 679 | 186.45 | 65.61 | 36.57 | 27.04 |
| BH1705_chrX | 661 | 179.71 | 64.16 | 35.89 | 26.56 |
| BH1706_chrX | 777 | 195.24 | 67.73 | 37.47 | 27.51 |
| BH1708_chrX | 684 | 209.37 | 93.17 | 58.22 | 44.41 |
| BH1713_chrX | 758 | 194.32 | 65.76 | 37.74 | 28.50 |
| BH1716_chrX | 764 | 201.74 | 69.01 | 38.30 | 28.30 |
| Mean | 725 | 194.54 | 71.80 | 41.25 | 30.71 |
| Std. Error | 15 | 3.10 | 2.98 | 2.34 | 1.88 |
| BH1701_chrY | 34 | 16.49 | 9.89 | 7.57 | 6.51 |
| BH1704_chrY | 32 | 16.02 | 8.95 | 6.49 | 5.49 |
| BH1705_chrY | 35 | 16.20 | 9.00 | 6.66 | 5.68 |
| BH1708_chrY | 35 | 16.03 | 8.95 | 6.70 | 5.76 |
| Mean | 34 | 16.185 | 9.198 | 6.855 | 5.860 |
| Std. Error | 0.707 | 0.110 | 0.231 | 0.243 | 0.224 |

**Table S3**. The mean (*per chromosome*) SNP alpha-diversity for each individual from the 9-individual cohort

| Subject | | $q=0$ | $q=1$ | $q=2$ | $q=3$ | $q=4$ |
|---|---|---|---|---|---|---|
| BH1696 | Mean | 906.04 | 305.62 | 146.28 | 92.99 | 70.67 |
| | Std. Err. | 94.68 | 35.43 | 18.92 | 12.22 | 9.17 |
| BH1700 | Mean | 905.83 | 305.95 | 147.97 | 94.06 | 71.32 |
| | Std. Err. | 94.45 | 36.02 | 19.53 | 12.60 | 9.45 |
| BH1701 | Mean | 900.33 | 304.07 | 145.78 | 92.16 | 69.76 |
| | Std. Err. | 93.68 | 35.20 | 18.61 | 11.90 | 8.90 |
| BH1704 | Mean | 902.21 | 304.22 | 144.86 | 91.30 | 69.03 |
| | Std. Err. | 93.84 | 35.20 | 18.62 | 11.83 | 8.76 |
| BH1705 | Mean | 902.42 | 302.15 | 144.33 | 92.24 | 70.60 |
| | Std. Err. | 94.59 | 34.81 | 18.44 | 11.98 | 9.05 |
| BH1706 | Mean | 906.21 | 309.00 | 150.30 | 96.60 | 73.84 |
| | Std. Err. | 94.80 | 35.97 | 19.72 | 13.09 | 9.94 |
| BH1708 | Mean | 905.04 | 306.91 | 148.50 | 95.09 | 72.65 |
| | Std. Err. | 94.66 | 35.10 | 18.72 | 12.15 | 9.17 |
| BH1713 | Mean | 906.29 | 302.33 | 145.29 | 93.50 | 71.72 |
| | Std. Err. | 94.42 | 35.15 | 19.37 | 13.04 | 10.03 |
| BH1716 | Mean | 904.79 | 304.81 | 146.40 | 93.36 | 71.03 |
| | Std. Err. | 94.35 | 35.08 | 18.74 | 12.13 | 9.07 |
| Mean | Mean | 904 | 305.006 | 146.635 | 93.478 | 71.180 |
| | Std. Error | 30 | 11.556 | 6.205 | 4.036 | 3.040 |

**Table S4**. The Wilcoxon significance tests for the differences between the healthy and diseased groups or between the male and female groups

| Treatments | | $q=0$ | $q=1$ | $q=2$ | $q=3$ | $q=4$ |
|---|---|---|---|---|---|---|
| Healthy *vs.* Disease | $\neq$ | 0.950 | 0.876 | 0.781 | 0.742 | 0.833 |
| | $>$ | 0.526 | 0.438 | 0.390 | 0.371 | 0.417 |
| | $<$ | 0.475 | 0.563 | 0.610 | 0.630 | 0.584 |
| Male *vs.* Female | $\neq$ | 0.871 | 0.891 | 0.898 | 0.889 | 0.922 |
| | $>$ | 0.436 | 0.446 | 0.449 | 0.445 | 0.461 |
| | $<$ | 0.565 | 0.555 | 0.552 | 0.556 | 0.540 |